# Acoustic signal detection through the cross-correlation method in experiments with different signal to noise ratio and reverberation conditions

S.Adrián-Martínez, M.Ardid*, M.Bou-Cabo, I.Felis, C.Llorens, J.A.Martínez-Mora, M.Saldaña

Universitat Politècnica de València, Institut d'Investigació per a la Gestió Integrada de Zones Costaneres (IGIC). C/Paranimf 1, 46730 Gandia, Spain

*mardid@fis.upv.es

**Abstract.** The study and application of signal detection techniques based on cross-correlation method for acoustic transient signals in noisy and reverberant environments are presented. These techniques are shown to provide high signal to noise ratio, good signal discernment from very close echoes and accurate detection of signal arrival time. The proposed methodology has been tested on real data collected in environments and conditions where its benefits can be shown. This work focuses on the acoustic detection applied to tasks of positioning in underwater structures and calibration such those as ANTARES and KM3NeT deep-sea neutrino telescopes, as well as, in particle detection through acoustic events for the COUPP/PICO detectors. Moreover, a method for obtaining the real amplitude of the signal in time (voltage) by using cross correlation has been developed and tested and is described in this work.

**Keywords:** Acoustic signal detection, cross-correlation method, processing techniques, positioning, underwater neutrino telescopes, particle detectors.

## 1 Introduction

Acoustic signal detection has become an object of interest due to its utility and applicability in fields such as particle detection, underwater communication, medical issues, etc. The group of Acoustics Applied to Astroparticle Detection from the Universitat Politècnica de València collaborates with the particle detectors ANTARES [1], KM3NeT [2] and COUPP/PICO [3]. Acoustic technologies and processing analyses are developed and studied for positioning, calibration and particle detection tasks of the detectors.

Acoustic emitters and receivers are used for the positioning systems of underwater neutrino telescopes ANTARES [4] and KM3NeT [5] in order to monitor the position of the optical detection modules of these telescopes. The position of optical sensors need to be monitored with 10 cm accuracy to be able to determine the trajectory of the muon produced after a neutrino interaction in the vicinity of the telescope from the Cherenkov light that it produces [6]. An important aspect of the acoustic positioning system is the time accuracy in the acoustic signal detection since the positions are evaluated from triangulation of the distances between emitters and receivers, which are determined from the travel time of the acoustic wave and the knowledge of the sound speed. The distances between emitters and receivers are of the order of 1 km. Therefore, the acoustic emitted signals suffer a considerable attenuation in the medium and arrive to the acoustic receivers with a low signal to noise ratio. The environmental noise may mask the signal making the detection and the accurate determination of its arrival time a difficult goal, especially for the larger future telescope KM3NeT with larger distances.

On the other hand, an acoustic test bench has been developed for understanding the acoustic processes occurred inside of the vessels of the COUPP Bubble Chamber detector when a particle interacts in the medium transferring a small amount of energy, but very localized, to the superheated media [7]. This interaction produces a bubble through the nucleation process. Under these circumstances the distance from the bubble to the vessel walls are very short (cm order) and a reverberant field generated by multiple reflections in the walls takes place. With these conditions, the distinction of the direct signal from reflection is quite difficult to achieve, being also quite complex to determine the time and amplitude of the acoustic signal produced.

The elaboration of protocols and post-processing techniques are necessary for the correct detection of the signals used in these tasks. Methods based on time and frequency analysis result insufficient in some cases. The first step consists of using the traditional technic of cross-correlation between the received signals and the emitted signals (expected) for localizing the source distance. In addition, the use of specific signals with wide band frequency or non-correlated such as sine sweep signals or Maximum Length Sequence (MLS) signal together with correlation methods increase the amplitude and the correlation peak narrows, this allows a better signal detection, improves the accuracy in the arrival time and the discernment of echoes.

In this work the detection of acoustic signals with a unique receiver under a reverberant field or a high noise environment is shown. The correlation method has been studied and applied for this purpose. Moreover, a method for obtaining the real amplitude of the signal (voltage) by using cross-correlation technique has been developed. Its validation has been done by comparing the results with the ones obtained by analytic methods in time and frequency domain, achieving a high reliability for the accurate detection of acoustic signals and the analysis of them. The results obtained in these tests in different environments using different kind of signals are shown.

In section 2 the cross-correlation technique is described, as well as the method proposed for signal detection. The application of the method under different situations: high reverberation, low signal-to-noise ratio (S/N) or very low S/N, is presented in section 3. Finally, the conclusions are summarized in section 4.

## 2      The cross-correlation method for signal detection

Cross-correlation (or cross-covariance) consists on the displaced dot product between two signals. It is often used to quantify the degree of similarity or interdependence between two signals [8]. In our case, since all measurements were recorded using digital acquisition systems, all signals under study have been evaluated in discrete time, so that the correlation between two signals *x* and *y* with the same *N* samples length is expressed by the following expression:

$$Corr\{x, y\}[n] = \sum_{m=1}^{N} x[m] \cdot y[m+n] \tag{1}$$

If we do *y* = *x* we obtain the autocorrelation of the signal *x*.

Figure 1 shows the appearance of the signals used in these studies: tones, sweeps, and MLSs. On top, there are these ideal signals in the time domain, that is, the generated signals by the electric signal generator equipment. In the middle row, the spectrum of each signal can be seen, where the different bandwidths can be appreciated. At the bottom, the autocorrelations of each signal show that the higher bandwidth signals have a narrower correlation peak, so, in principle, they are easier to detect. To understand the importance and convenience of using these signals in each detector, the reader can look at articles [9,10].

It is worth to note that, in the cases shown, the correlation peak amplitude ($V_{max,corr}$) is the same and equal to the number of samples of the signal in question (*N*). Therefore, it can be obtained the peak voltage of the signal ($V_p$) by the following expression:

$$V_p = \frac{2 V_{max,corr}}{N} \tag{2}$$

Furthermore, this ratio does not vary with the amplitude of the signal and is less susceptible to the presence of noise.

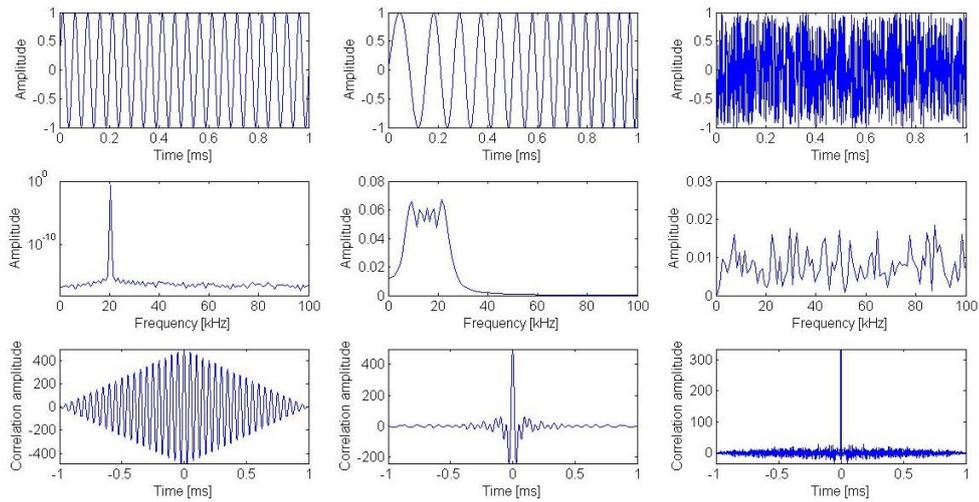

**Fig. 1.** Signals used for acoustic studies: tone, sweep and MLS.

However, the interest is the use of the method for the accurate detection of signals and the recorded signals will be influenced by reflections and noise that may vary the amplitude and profile of the direct signal detection.

Figure 2 shows the case of a tone, a sweep and MLS received signals with a distance of 112.5 m between emission and reception (E-R). On the top, the receiving signals in time domain after applying a high order band pass filter are shown (the original recorded signal in time is so noisy that the receiving signal is completely masked). On the bottom, it can be seen the cross-correlation of each signal (without prefiltering) where direct signal reflections are easier and more effective to discern that working in the time or frequency domains, especially for high bandwidth signals (narrower auto-correlation peak).

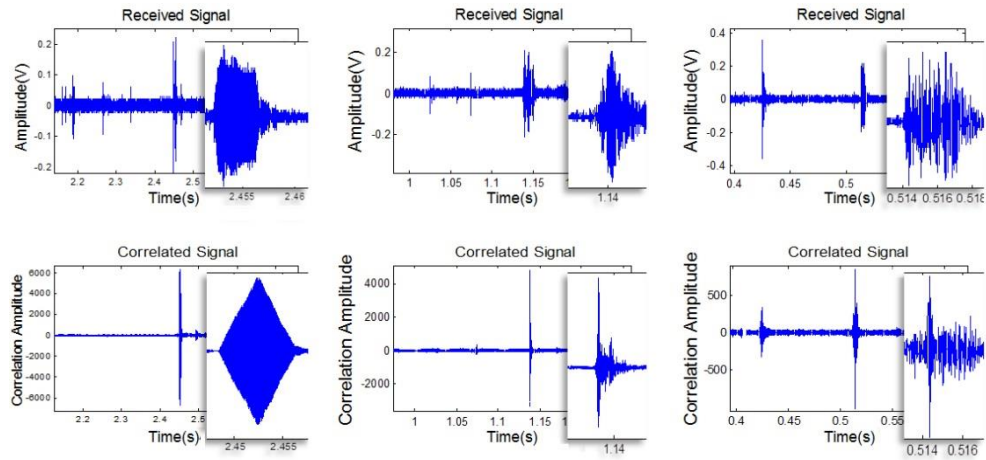

**Fig. 2.** Example of recorded signals at 112.5 m Emitter-Receiver distance in the harbour of Gandia.

Nevertheless, using this it is only possible to locate the signal but cannot know a priori the peak amplitude of the signal. This is because trying to tackle the problem from both time and frequency domains is completely crucial windowing temporarily the direct signal avoiding reflections to obtain a reliable value of its amplitude, which is not always possible.

Then, it would be important to obtain the corresponding relation between the maximum of the cross-correlation between received and emitted signal with the amplitude of the received signal avoiding reflections. This issue has been studied and has been found that if the amplitude of the signal sent ($V_{p,env}$), its number of samples ($n_{env}$) and the maximum correlation value ($V_{max,corr}$) corresponding to the detection of this signal are known, then it is possible to obtain the peak-amplitude voltage of the received signal applying the following expression:

$$V_{p,rec} = \frac{V_{\max,corr}}{V_{p,env}} \frac{2}{n_{env}} \qquad (3)$$

In the following sections the results of applying this equation to the results of the correlations obtained and compared with values obtained applying time and frequency domain methods are presented. In addition, the improvements obtained by using this technique in terms of detection accuracy in different acoustical environments are also shown.

## 3 Application

The different conditions in which the measurements of acoustic detection were performed are: inside a small vessel, in a tank of acoustic test, in a pool, in the harbour of Gandia, and in ANTARES deep sea neutrino telescope. Although under different conditions of pressure, salinity and temperature, the acoustic propagation media in all the tests is water. Table 1 shows the relationship between the wavelength range associated with the studied signals ($\lambda$) and the geometrical dimensions of the places where acoustic processes occur ($l$).

| *Measure condition* | *Characteristic distance l [m]* | *$\lambda / l$* |
|---|---|---|
| *Vessel* | *0.02* | *2.2* |
| *Tank* | *0.05-1* | *0.22* |
| *Pool* | *4* | *0.022* |
| *Harbor* | *120* | *0.0005* |
| *Sea* | *200* | *0.0003* |

**Table 1.** Characteristics of the acoustic conditions of the different measurements and tests.

With this, it follows that conditions with higher ratio $\lambda/l$ means working in a reverberant field, with a higher complexity, while configurations with a smaller $\lambda/l$ ratio means that there is a less reverberant field, but usually a lower S/N ratio. As discussed below, both extreme situations make difficult the process of acoustic detection.

The results obtained in these conditions, the acoustic systems used in transmission and reception, and the results in terms of improvement of signal detection and S/N using cross-correlation method are shown in the following sections.

### 3.1 High reverberation conditions

When emitter and receiver are close and the dimensions of the enclosure where the acoustic processes occur are comparatively small, both signal and reverberation are high. This is the case of the configurations shown in Figure 3 that corresponds to a part of the acoustic test bench for COUPP detector [11]. On the left, the two experimental setups are shown. The first one corresponds to acoustic propagation studies inside a vessel, and the second one was used to study the acoustic attenuation. On the right the transducers used are shown. The signal was emitted with the pre-amplified ITC 1042 transducer and received with the needle-like RESON TC 4038 transducer.

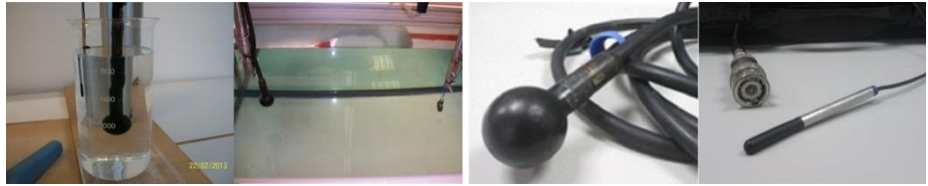

**Fig. 3.** Experimental setups (left) and transducers used (right).

Figure 4 shows an example of a 30 kHz tone of 5 cycles of duration emitted and recorded under these conditions and their cross-correlation. It can be seen that the maximum of the correlation corresponds with the reception time of the received signal.

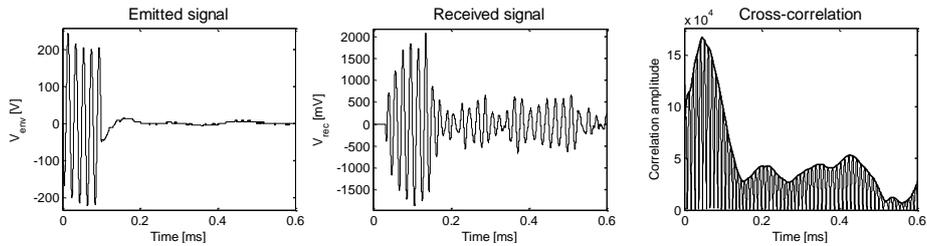

**Fig. 4.** Example of emitted signal, received signal, and cross-correlation.

Figure 5-left shows that for the tones studied between 10 kHz and 100 kHz the accuracy of this method is quite good, with an error smaller than 10 %. Considering the characteristic dimensions of the problem and 1500 m/s as sound propagation speed, this uncertainly is of the same order of magnitude of the experimental uncertainly (1 mm). As expected, the maximum deviation corresponds to lower frequencies, and it seems there is some frequency dependent fluctuations. This can be another argument in favour of using broadband signals for cross-correlation techniques.

The received amplitudes of the signals have been obtained using equation (3). The results are shown in Figure 5-right compared to the results obtained with standard techniques in time and frequency domains. It can be observed that the results are very similar.

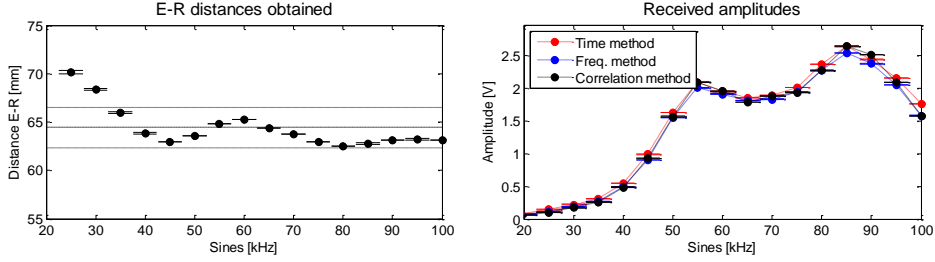

**Fig. 5.** Left: distances obtained between emitter and receiver by cross-correlation with tones between 20 kHz to 100 kHz. Right: Received amplitudes through the cross-correlation method using Eq.3 and using time and frequency domain methods.

### 3.2 Low signal to noise ratio conditions

The following configuration used is an intermediate step between high signal to noise ratio (section 3.1) and very low signal to noise (measurements in the ANTARES neutrino telescope, presented in the next section 3.3). This is the case of measurements taken on a pool as shown in figure 6 (left). In this experimental setup, the transmitter consists of an array of three transducers FFR SX83 (middle) and an electronic board to generate and amplify the different acoustic signals. This system can operate in three different modes: emitting with a single element, with the three elements connected in series and the three elements connected in parallel [10]. Our measures were made with the transducers connected in parallel so, in this embodiment, higher transmission power is obtained. The reception was performed using a FFR SX30 (right).

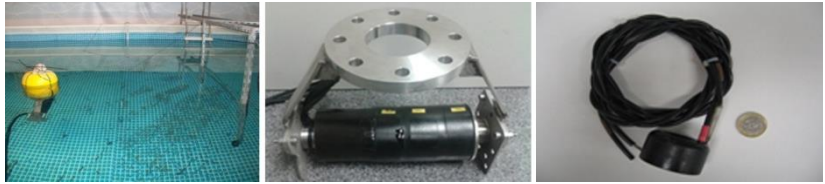

**Fig. 6.** Experimental setup (left), emitter (middle) and receiver (right) transducers.

Using tones between 10 kHz and 60 kHz in these conditions, we have calculated the emitter-receiver distances from flight times, as described above. The results are shown in figure 7 and compared with those obtained directly in time-domain method. In this case, we can see that the deviation of the measurements relative to a mean value is 5%, which corresponds to an uncertainty less than ± 20 cm. However, if we discard some out-layer measure (sine of 40 kHz) the deviation of the values is reduced to 2.3%, i.e., ± 9 cm. We think that a reason for the relatively large variation between different measurements at different frequencies might be the interference between the three emitters of the array, which depends on the frequency. Again here, the use of broadband signals with the cross-correlation method may help to mitigate this problem since it will average the response of the different frequencies.

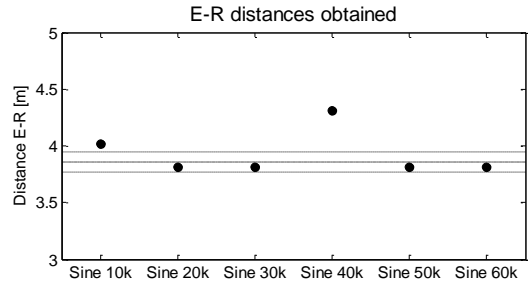

**Fig. 7.** Emitter-receiver distances obtained by cross-correlation method using tones between 10 kHz to 60 kHz (considering 1500 m/s as the sound propagation speed).

The plots of figure 8 show the results obtained by comparing the voltages (left) and the S/N ratios (right) both in cross-correlation method and time-domain method (in this case, since the signals can be windowed properly, avoiding the presence of reflections, values obtained in time and frequency domains are coincident).

As before, using the Eq. 3 very similar results to the usual techniques are obtained. On the other hand, the S/N ratio increases considerably (at least 20 dB) for the set of signals used using correlation method. This improvement is crucial for a correct detection of the signals.

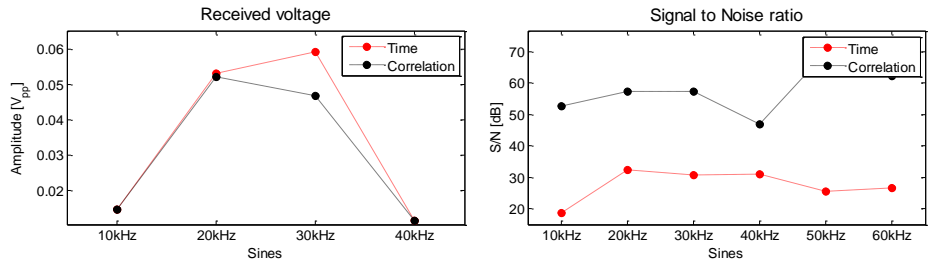

**Fig. 8.** Comparison of cross-correlation and time domain method to obtain the received voltage amplitude (left) and the S/N ratio (right).

### 3.3 Very low signal to noise ratio conditions

The more complex environment in which this study has been performed is the acoustic measurements made in situ in deep-sea at the ANTARES site. In this case, the distance between emitter and receiver was about 180 m and the S/N ratio was quite low. Figure 9 shows on the left an artistic and schematic view of the telescope. The emitter was a FFR SX30 transducer, shown in the middle, with an electronic board designed specifically for this type of transducer to optimize and amplify the signal sent [10], and the receiving hydrophone was a HTI-08 transducer, shown on the right [12].

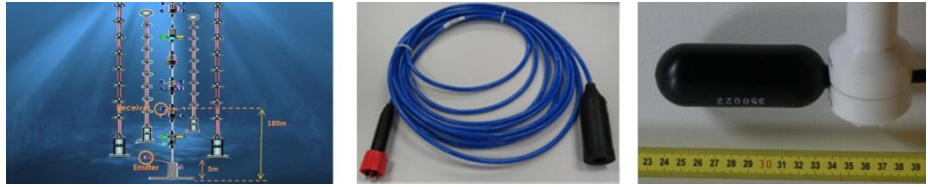

**Fig. 9.** View of the ANTARES neutrino telescope (left) and pictures of the emitter FFR SX30 (middle) and of the receiver HTI-08 (right) transducers.

Since in this ANTARES test synchronization between transmitter and receiver was not available, it is not possible to calculate absolute flight times. However, the received amplitudes expression as well as the increase of the S/N ratio obtained by cross-correlation method can be evaluated here, as shown in figure 10. In this case, sine signals of 20, 30 and 40 kHz, sweep signals between 20 to 48 kHz, and 28 to 44 kHz, and MLS signals were used.

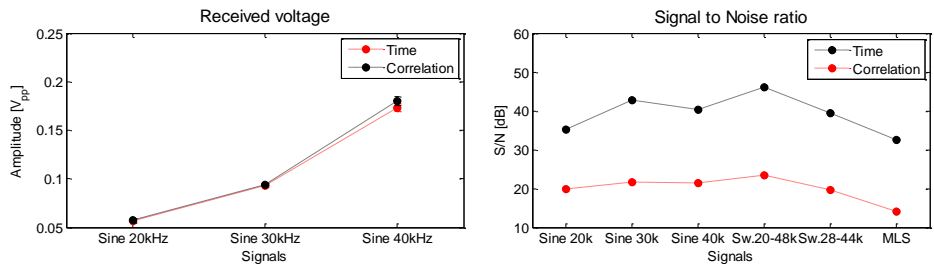

**Fig. 10.** Received amplitude (left) and S/N ratio (right) both in cross-correlation and time domain method.

It can be concluded from these measurements that using the cross-correlation method is possible to obtain the signal amplitude accurately and obtain an increase of 15 to 20 dB in the S/N ratio, with a consequent improvement in the acoustic detection.

Additionally, and with the aim of applying this technique for post-processing signals in the future KM3NeT neutrino telescope, simulations of propagation of signals measured in ANTARES over longer distances have been done. Figure 11 shows, the improvement in the S/N ratio as a function of the distance using the different signals and methods.

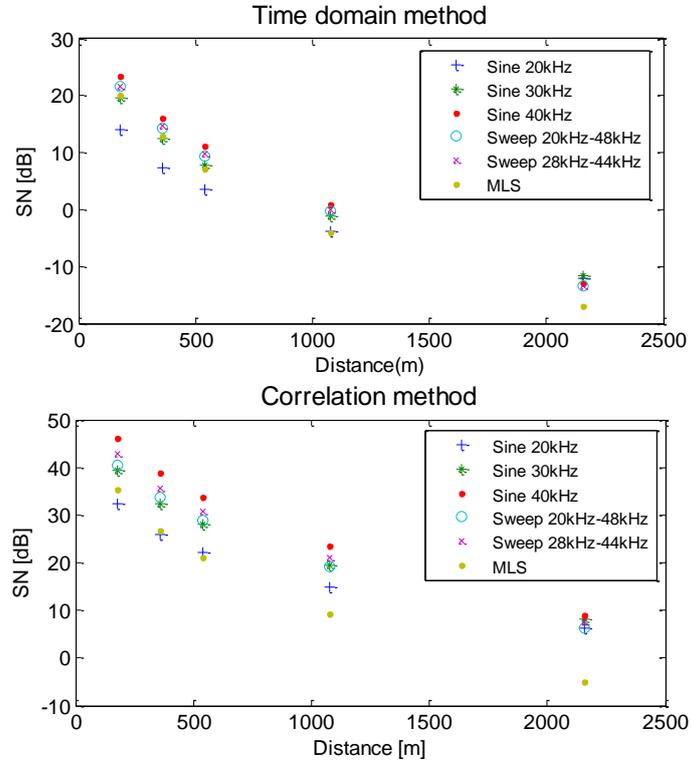

**Fig. 11.** S/N ratio obtained using time domain method (top) and cross-correlation method (bottom).

## 4 Conclusions

We have seen that, using different signal emission-acquisition systems, working on a wide range of distances and in very different environmental conditions, good acoustic detection through the technique of cross-correlation between the emitted and received signals can be obtained. This technique is more favourable for broadband signals (sweeps and MLS) because they have a narrower correlation peak and consequently they are easier to discern than others peaks. Furthermore, this technique is powerful in measurement conditions with a reduced S/N ratio, as the case in marine environments over long distances where the recorded signal is weak, or in environments with high background noise. In addition, we have obtained a relation between the peak value of the cross-correlation and the voltage value of the received signal, which synthesizes and optimizes the signal analysis.

## Acknowledgements

This work has been supported by the Ministerio de Economía y Competitividad (Spanish Government), project ref. FPA2012-37528-C02-02, Multidark (CSD2009-00064). It has also being funded by Generalitat Valenciana, Prometeo/2009/26. Thanks to the ANTARES Collaboration for the help in the measurements made in the ANTARES deep-sea neutrino telescope.